\documentstyle[prd,aps]{revtex}
\begin{document}
\draft
\title{Does a Kalb--Ramond field make spacetime optically active ?} 
\author{Sayan Kar \footnote{Electronic address: {\em sayan@cts.iitkgp.ernet.in}}
${}^{(1)}$, Parthasarathi Majumdar\footnote{Electronic address: 
{\em partha@imsc.ernet.in
}}${}^{(2)}$Soumitra SenGupta\footnote{Electronic address: {\em 
soumitra@juphys.ernet.in}}${}^{(3)}$
and Aninda Sinha\footnote{Electronic Address : {\em as402@cam.ac.uk}}${}^{(4)}$}
\address{{\rm $^{(1)}$}Department of Physics and Centre for
Theoretical Studies \\Indian Institute of Technology, Kharagpur 721 302, India}
\address{{\rm $^{(2)}$} The Institute of Mathematical Sciences, \\
CIT Campus, Taramani, Chennai 600 113, India} 
\address{{\rm $^{(3)}$}Department of Physics 
, Jadavpur University, Calcutta  700 032, India}
\address{{{\rm $^{(4)}$} St. Edmunds College, Cambridge CB3 0BN, University
of Cambridge, UK}} 

\maketitle

\begin{abstract}
{A spacetime with torsion produced by a Kalb-Ramond field coupled 
gravitationally to the Maxwell field, in accordance with a recent proposal 
by two of us (PM and SS), is argued to 
lead to an optical activity in 
synchrotron radiation from cosmologically
distant radio sources. We suggest that
this could {\em qualitatively} explain   
observational data from a large number
of radio sources displaying such
polarization asymmetry (after eliminating effects of Faraday rotation due
to magnetized galactic plasma). 
Possible implications for heterotic string theory
are also outlined. 
}
\end{abstract}


\maketitle

\section{Introduction}

The massless antisymmetric tensor Kalb-Ramond (KR) field has been an inherent
aspect of supergravity theories. Indeed, the tensor multiplet in $N=1$ theories
has interesting duality properties that are exploited in its coupling to
supergravity \cite{ggrw}. In extended supergravity, the KR field becomes a part
of the supergravity multiplet itself, thus playing a more intrinsic role. The
importance of the KR field in supergravity theories in various spacetime
dimensions has been emphasized more than ever in string theories \cite{gsw}. 
Supergravity multiplets constitute the massless sector of string theories. As
such, they inevitably contain massless KR fields. Such fields implement a
spacetime background for string theory possessing {\it torsion} in addition to
curvature. 

One aspect which particularly deserves attention in this respect is that of
cosmology in the presence of torsion, or, equivalently, in the presence of a KR
field.  Indeed, the cosmological domain is the most likely arena for physical
`stringy' effects to appear. Since sources of torsion exist in the massless
spectra of most viable string theories (in the form of the KR field), at least
in the perturbative sector, cosmological models with non--zero torsion need to
be investigated. Restricting ones attention to Einstein-Cartan spacetimes, the
issue of gauge invariant coupling to standard massless gauge fields arises. The
well-known problem \cite{ham} associated with the Maxwell field has been
addressed in \cite{pmss} by introducing a KR field, and augmenting it in accord 
with requirements of quantum consistency of heterotic string theory toroidally 
compactified to four spacetime dimensions. 

In this paper, we examine the consequences of the resulting dynamics, to discern
effects that could, even if remotely, be astrophysically/cosmologically
observable. The possibility that a KR field may induce a rotation of 
the plane of polarization of electromagnetic radiation from
cosmologically distant sources, was already alluded to in \cite{pmss}. There is 
some evidence that optical activity of a related type may have already been 
{\it observed} in radiation from distant quasars and
other radio sources \cite{jr}.\footnote{Earlier analyses of the data on this topic
\cite{nr} (and the data itself) were infused with a certain amount of controversy
\cite{crit}; later works appear to be free of most of the contentious issues 
\cite{jain}.} Typically, the observed angle of rotation of the plane of
polarization can be expressed as {\cite{nr}} :
\begin{equation}
\theta = \alpha \lambda^{2} + \chi \label{rot}
\end{equation}
where $\alpha$ (the Faraday rotation measure) and $\chi$ are constants and $\lambda $ is the wavelength
of the electromagnetic wave. $\chi$ is the angle between a reference axis
and the electric field of the wave  
when it is emitted from the source 
galaxy, while the first term, by dint of its dependence on the wavelength 
quadratically, represents {\it Faraday} rotation of the plane of 
polarization due to passage of the electromagnetic wave through galactic (and 
possibly inter-galactic) magnetized plasmas. The key aim of this paper is to 
argue that Einstein-Kalb--Ramond--Maxwell coupling can be responsible, albeit 
qualitatively, in explaining the origin of this 
extra
 bit of rotation ($\chi$ ). More precisely, the question is
whether the observed
values of $\chi$ can be explained by {\em assuming} that the plane of
polarisation of the wave emitted is initially along a fixed angle
(0 or $\frac{\pi}{2}$, in accordance with the source models for
elliptical galaxies and the nature of synchrotron radiation) relative to the galaxy major axis (note that the major axis is tilted w.r.t. the reference
axis by an angle $\psi$ known as the intrinsic position angle of the
corresponding galaxy), but undergoes a rotation due to
the presence of the KR field. 
We recall that there is no extant proposal based on fundamental physics 
to account for $\chi$. 

A result in support of this proposal is most likely an
evidence of the existence of a primordial KR field and hence of a spacetime with
(non-propagating) torsion in an early epoch. Since a KR field does occur naturally in 
some supergravity theories and hence in the 
massless spectrum of closed string theory, such an observation may
perhaps be construed to be evidence for supergravity as well as being a 
hint of an underlying string structure. 

A remark on our basic strategy is perhaps in order: in the sequel, we treat the KR
field as a {\it tiny} perturbation on the Maxwell field equations in a standard
cosmological background. In other words, despite the assumed `primordial' origin of
the KR field, we restrict its strength to such small values that its energy density
plays an insignificant role in shaping geometry on a cosmological scale.  For the
latter, we choose two standard scenarios, viz., the {\it spatially} flat
Friedmann-Robertson-Walker background with the scale factor, which depends only on
(comoving) time, evolving according to both a radiation dominated and a matter
dominated scheme. This is, of course, an approximation which we hope to improve
in future assays on this subject. It is primarily motivated by the fact that there
may not exist exact solutions of the KR-coupled Einstein equation for the evolution
of spacetime geometry, which are also homogeneous {\it and} isotropic. There is no
compelling observational evidence yet to doubt these latter requirements. 

The paper is organized as follows: in Section II, we review the important aspects of
the earlier paper \cite{pmss} as background for the present work. This is followed
in Section III by a presentation of the solution of Maxwell-KR field equations in a
flat background spacetime and retaining only the leading order coupling of the two
fields. The first hint of an optical activity
 already appears at this preliminary stage.
Sections IV constitutes the main part of the work, where, in a spatially flat
FRW background, the Maxwell-KR equations are considered for conformal
factors pertaining to the radiation and matter dominated scenarios. 
Expressions for the angle of rotation of the plane of polarization are obtained,as a function of the redshift, for very large co-moving times, both
in the radiation and matter dominated cosmological settings. 
We conclude in Section V. 

\section{Einstein-Maxwell-Kalb-Ramond Coupling} 

Let us briefly recapitulate the main tenets of the earlier paper \cite{pmss}.
It is well known that the electromagnetic field tensor, defined as the 
generally
covariant curl of the four potential, is not invariant under  the standard
$U(1)$ electromagnetic gauge transformation $\delta A_{\mu} =
\partial_{\mu} \omega$, assuming that the torsion tensor
$T^{\rho}_{\mu \nu}$ - a purely geometric quantity like curvature must be 
gauge invariant. We introduce a Kalb-Ramond (KR)
antisymmetric second rank tensor field $B_{\mu \nu}$ as a possible
source of torsion. The KR field strength is  modified by $U(1)$ Chern-Simons
terms which originates from the quantum consistency of an underlying 
string theory \cite{gsw}. This augmented field is coupled to the torsion 
in a way that the
resulting action preserves all gauge symmetries and has the form 
\begin{eqnarray}
S ~&=&~ \int d^{4}x~ \sqrt{-g}~[~{1 \over 16\pi G}~ R~(g~,~T)~-~\frac14 F_{\mu \nu}
~F^{\mu
\nu}~-\frac12~{\tilde H}_{\mu \nu \lambda} {\tilde H}^{\mu \nu \lambda}~\nonumber \\
&+&~{1 \over \sqrt{G}}~T^{\mu \nu \lambda}~{\tilde H}_{\mu \nu \lambda}~ ]~\label{act}
\end{eqnarray}
where $R$ is the scalar curvature, defined as 
$R = R_{\alpha\mu\beta\nu}g^{\alpha\beta}g^{\mu\nu}$. $R_{\alpha\mu\beta\nu}$ 
is the Riemann-Christoffel tensor:
\begin{equation}
R^{\kappa}_{\mu\nu\lambda} =
\partial_{\mu} \Gamma^{\kappa}_{{\nu\lambda}}
- \partial_{\nu} \Gamma^{\kappa}_{{\mu\lambda}} + 
\Gamma^{\kappa}_{\mu\sigma} \Gamma^{\sigma}_{\nu\lambda} - 
 \Gamma^{\kappa}_{\nu\sigma} \Gamma^{\sigma}_{\mu\lambda}  
\end{equation}    
The torsion tensor $T_{\mu \nu \lambda}$ is an auxiliary field in eq.
(\ref{act}),
obeying the constraint equation
\begin{equation}
T_{\mu \nu \lambda}~~=~~\sqrt{G}~{\tilde H}_{\mu \nu \lambda}~, \label{tors}
\end{equation}
where, ${\tilde H}_{\mu \nu \lambda}~\equiv~\partial_{[\mu}B_{\nu
\lambda]} + \frac13 \sqrt{G} A_{[\mu} F_{\nu \lambda]}$ \cite{pmss}.  Thus, the
augmented KR field strength three tensor plays the role of the
spin angular momentum density which is the source of torsion \cite{Hehl}.
Substituting the above equation in (\ref{act}) and varying with respect to
$B_{\mu \nu}$ and $A_{\mu}$ respectively, we obtain the equations
\begin{equation} 
D_{\mu}~{\tilde H}^{\mu \nu \lambda}~~=~~0 \label{bmn}
\end{equation}
and
\begin{equation}
D_{\mu}~F^{\mu \nu}~~=~~\sqrt{G}~{\tilde H}^{\mu \nu \lambda}~ F_{\lambda \mu}~.
\label{max}
\end{equation}

Now, the KR three tensor is Hodge-dual to the derivative of a spinless field
$\phi$, so that, after a partial integration, one obtains,
\begin{equation}
S_{int}~=~\frac12~\int d^4 x ~\phi~F_{\mu \nu}~^*F^{\mu \nu}~,
\label{inter}
\end{equation}
where, $^*F^{\mu \nu}~\equiv~\epsilon^{\mu \nu \lambda \sigma}F_{\lambda
\sigma}$. Here, we have noted the fact that 
\begin{equation}
{1 \over \sqrt{-g}} \partial_{\nu}~(\sqrt{-g} ~^*F^{\mu
\nu})~=~D^S_{\mu}~^*F^{\mu \nu}~=~0
\end{equation}
by the Maxwell Bianchi identity, where, $D^S$ is the covariant derivative 
using the Christoffel connection.

In general, in addition to the graviton and the KR field, the perturbative 
sector of the heterotic string contains a scalar dilaton field whose dynamics
is also known to have cosmological consequences \cite{ven}. In this paper, 
we shall however ignore this dynamics for the moment and focus instead on what
the KR field does. In any event, the dilaton field couples to the Maxwell
Lagrangian and the kinetic term of the KR field, and so cannot affect in any 
major way the optical activity induced by the axion (KR) field; the latter 
effects appear due to the fact that $F ^*F$ is {\it pseudoscalar}. Freezing 
the dilaton field and taking the KR field strength to be 
$$H_{\mu\nu\rho}~=~\epsilon_{\mu\nu \rho}^{~~~\sigma} D_{\sigma}H~, $$
where $H$ is a  pseudoscalar, one obtains the modified generally covariant
Maxwell equations \cite{pmss}
\begin{eqnarray} 
{\bf D} \cdot {\bf E}~&=&~2 \sqrt{G} {\bf D}H \cdot {\bf B} \nonumber \\
D_0{\bf E}~-~{\bf D} \times {\bf B}~&=&~-2 \sqrt{G} [ D_0 H {\bf B}~-~{\bf
D}H \times {\bf E}] \nonumber \\
&+&~2G[({\bf B}^2~-~{\bf E}^2) {\bf A}~+~({\bf A} \cdot {\bf E}){\bf
E}~+~({\bf A} \cdot {\bf B}){\bf B}]~\nonumber \\
D_0{\bf B}~+~{\bf D} \times {\bf E}~&=&~0~=~{\bf D} \cdot {\bf B} ~.
\end{eqnarray}
Here $D_{\mu}$ is the covariant derivative in the spatially flat FRW metric. To a 
first approximation, we drop
the $O(G)$ terms arising from the stringy augmentation of the KR field strength 
in terms of the Chern Simons three form. Next, redefine the pseudoscalar 
field $H$ to absorb the $\sqrt{G}$, so that this field becomes dimensionless. 
The equations now look like
\begin{eqnarray} 
{\bf D} \cdot {\bf E}~&=&~2 {\bf D}H \cdot {\bf B} \nonumber \\
D_0{\bf E}~-~{\bf D} \times {\bf B}~&=&~-2  [ D_0 H {\bf B}~-~{\bf
D}H \times {\bf E}] \nonumber \\
D_0{\bf B}~+~{\bf D} \times {\bf E}~&=&~0~=~{\bf D} \cdot {\bf B}
~.\label{maxx}
\end{eqnarray}
The last equations in the array constitute the Maxwell Bianchi identity. 
In a spatially flat isotropic FRW background with metric
\begin{equation}
ds^2 ~=~R^2(\eta)~(d\eta^2~-~d {\bf x}^2)~, \label{metr}
\end{equation}
where, $\eta$ is the conformal time coordinate, defined by $d\eta =
dt/R(t)$, the above equations assume the form,
\begin{eqnarray} 
{\bf \nabla} \cdot {\bf E}~R^2~&=&~2 {\bf \nabla}H \cdot {\bf
B}~R^2 \nonumber \\
\partial_{\eta}({\bf E}~R^2)~-~{\bf \nabla} \times {\bf B}~R^2~&=&~-2  [
\partial_{\eta} H {\bf B}~R^2~-~{\bf
\nabla}H \times {\bf E}~R^2] \nonumber \\
\partial_{\eta}({\bf B}~R^2)~+~{\bf \nabla} \times {\bf
E}~R^2~&=&~0~=~{\bf
\nabla} \cdot {\bf B}~R^2 ~.~\label{max2}
\end{eqnarray}

\section{Flat universe}

We first consider the simple situation corresponding to a {\it flat} background
spacetime ($R(\eta)=1$), just to obtain a preliminary understanding of the
effects involved. This simplification does not in any way reduce the
qualitative aspects of the optical effects under discussion, although the
quantitative details obtained in this manner may not be reliable. Recall that the 
KR field strength $H_{\mu \nu \rho} =
\partial_{[\mu} B_{\nu \rho]}$, so that it satisfies the Bianchi identity 
\begin{equation}
\epsilon^{\mu \nu \lambda \sigma} \partial_{\sigma} H_{\mu \nu
\lambda}~=~0~. \label{hbi}
\end{equation}
This immediately implies that the pseudoscalar $H$ satisfies the massless
Klein-Gordon eqn $\Box H = 0$. For non-flat 
backgrounds, the d' Alembertian operator is to be replaced by its generally 
covariant counterpart. We note that this is a departure from approaches in the 
literature where the axion (KR) field $H$ is introduced ad hoc with no specified
dynamics; here the Bianchi identity of the dual KR field is precisely the
equation of motion of the axion. Assume now that $H$ is only a function of 
the comoving time
coordinate $\eta$, so that, the Klein-Gordon equation reduces to the simple equation
${d^2 H \over d \eta^2} =0$ with the obvious solution $H = h \eta + h_0$,
where $h$ and $h_0$ are constants. This spatial homogeneity of the Klein-Gordon 
field is possibly a justified assumption over the cosmologically long distance scales
of our interest.  

Proceeding along the lines of \cite{cf} and \cite{cfj},  we arrive at the
equation
\begin{equation}
{d^2 b_{\pm} \over d \eta^2}~+~(k^2~\mp~2~h~k)~b_{\pm}~=~0~,
\label{pol}~\end{equation}
where we have decomposed ${\bf B} = {\bf b}(\eta)~e^{i{\bf k} \cdot {\bf
x}}$ and have chosen the $z$ direction to be the propagation direction
of the electromagnetic wave. The circular polarization states are defined as
$b_{\pm} \equiv b_x \pm i b_y$. Unlike the corresponding
equation in \cite{cf}, eqn (\ref{pol}) can be solved {\it exactly} :
\begin{equation} 
b_{\pm}~~=~~b_0~e^{i \omega_{\pm}~\eta} ~\equiv b_0 e^{i\phi_{\pm}},
\label{sol} \end{equation}
where, $\omega_{\pm}^2 \equiv k(k \mp 2h)$. 
The optical activity due to the presence of the KR field is 
thus given by the difference
\begin{equation}
\left ( \Delta \phi \right )_{mag}\equiv \frac12 (\phi_+ - \phi_-) =  -~h~\eta \hspace{.2in}
for 
\hspace{.1in} k
\gg h
\end{equation} 

We also note that the equation for the electric field (with an assumption
${\bf E} = {\bf e(\eta)} e^{ikz}$ and a similar definition for
$e_{\pm}$) takes the form :

\begin{equation}
\frac{d^{2}e_{\pm}}{d\eta^2} + k^2 e_{\pm} = -2h\frac{db_{\pm}}{d\eta}
\end{equation}

A solution of the electric field equation is therefore dependent on the
solution of the magnetic field equation. It is easy to see that, 

\begin{equation}
e_{\pm} = \mp \frac{b_0}{k}\omega_{\pm} i e^{i\omega_{\pm} \eta}
\end{equation}

is a solution for the electric equation and the amount of rotation is
the same for both the electric and magnetic fields.
  
All this is clear indication of an optical activity induced by the KR field.
It is quite unlikely that this effect will disappear 
when cosmological backgrounds
with curvature, the dilaton field or indeed the Chern Simons terms are included in
the Maxwell equations; these features must be incorporated before a detailed
comparison with any observational result can be made. 

\section{Spatially flat FRW universe: radiation and matter dominated cases}

The next immediate step in our analysis is to solve Maxwell equations once
again, but in a non-trivial cosmology - we choose for simplicity the {\it
spatially} flat Friedman--Robertson--Walker (FRW) type of background. 

The equation of motion of the 
pseudoscalar field is given as :

\begin{equation}
\Box  H = 0
\end{equation}

where $\Box$ is now the covariant d'Alembertian appropriate to the
spatially flat metric.

For a spatially independent H field, such that $H\equiv H(\eta)$
(19) has a first integral of the
form 
\begin{equation}
\partial_0 {H} = \frac{h}{R^{2}(\eta)}
\end{equation}
where h is an integration constant, which, in a sense, is a `measure' of
the pseudoscalar $H$ field or, equivalently, the dual three form field 
$H_{\mu\nu\lambda}$. 

The equations that the polarization states $b_{\pm}$ satisfy for such a 
background can be similarly written down
in terms of the quantity $F_{\pm}$ where $b_{\pm} = \frac{F_{\pm}}{R^{2}}$.
They are 
\begin{equation}
{d^2 F_{\pm} \over d \eta^2}~+~(k^2~\mp~{2~h~k\over
R^2(\eta)})~F_{\pm}~=~0~,
\label{pol2}~\end{equation}

As for the flat spacetime case, a corresponding equation for the
electric field polarisation states $e_{\pm}$ can also be obtained
in terms of a quantity $G_{\pm}$ where $e_{\pm}=\frac{G_{\pm}}{R^{2}}$
. This turns out to be :

\begin{equation}
\left ( \frac{d^{2}}{d\eta^2} + k^2 \right ) G_{\pm} = -2 h
\frac{d}{d\eta} \left [ \frac{F_{\pm}}{R^{2}} \right ]
\end{equation}

Note, that the electric field equations are dependent on the 
solution of their magnetic field counterparts. The magnetic
field equations, however, can be solved without any reference to
the electric ones. 
The rotation can
indeed be calculated for both the electric and magnetic fields and
we demonstrate our results for the magnetic field case in the
sections below.

The equations, of course, can be solved explicitly only after a  
knowledge of the scale factor $R(\eta)$ is available. One may resort to a WKB 
approximation along the lines of {\cite{cf}} and arrive at qualitative results.  
We prefer, as an alternative exercise, to choose a `physically reasonable' 
scale factor and derive the  effects of optical activity for such a case.  As 
mentioned earlier, the actual 
situation will correspond to a scale factor  corresponding to a solution of the
Einstein-KR-Maxwell-dilaton equations of motion for low energy
effective supergravity. However, since the dilaton couples
to the Maxwell Lagrange density and not to $F~^*F$, it is unlikely
that the effect that we find without it will be washed away by its 
inclusion.

Let us assume a scale factor $R(\eta) = \eta/\eta_0^R$ which is
equivalent, in real time, to the scale factor of a radiation dominated FRW
model, for $1/\eta_0^R=\left(8\pi G\epsilon_0/3 \right)^{1/2}$, with
$\epsilon_0$ being the primordial radiant energy density. For a
matter dominated model we assume $R(\eta) = \left ( \eta/\eta_0^M \right )^2$.
Our
objective is to obtain the {\it asymptotic} dependence on $\eta$ and
the parameters of the theory 
which are 
$\tilde h = h(\eta_0^R)^2$ , $h'=h(\eta_0^M)^4$ and the wave number $k$.

Accordingly, we have the two equations for the 
radiation and matter dominated cases which we quote below.
 
\begin{equation}
{d^2F_{\pm} \over dx^2} +\left (1 - {\mu_{\pm}^2 \over x^2}
\right) F_{\pm} ~=~0 ~,~\mu_{\pm}^2 \equiv 2{\tilde h}k~\label{rd}
\end{equation}
and 
\begin{equation}
{d^2F_{\pm} \over dx^2} +\left (1 - {\mu_{\pm}^2 \over x^4}
\right) F_{\pm} ~=~0 ~,~\mu_{\pm}^2 \equiv 2h'k^3~\label{md}
\end{equation}
In the above, we use dimensionless quantities throughout, with
$x=k\eta$

We now use the ansatz
\begin{equation}
F_{\pm}(x) = e^{ix}~v_{\pm}(x)
\end{equation}
so that, (\ref{rd}) and (\ref{md}) reduce to
\begin{eqnarray}
{d^2v_{\pm} \over dx^2} + 2i {dv_{\pm} \over dx} - {\mu_{\pm}^2 \over x^2}
v_{\pm} &=& 0 ~\label{rad} \\
{d^2v_{\pm} \over dx^2} + 2i {dv_{\pm} \over dx} - {\mu_{\pm}^2 \over x^4}
v_{\pm} &=& 0 ~\label{mad}
\end{eqnarray}

We are only interested in asymptotic solution of these equations for 
$x \rightarrow + \infty$. Accordingly, we choose a solution for both cases
of the type
\begin{equation}
v_{\pm}(x) = v_0^{\pm} + {v_1^{\pm} \over x} + {v_2^{\pm} \over x^2} +
\cdots
\end{equation}
This is an asymptotic solution, as given standard texts on differential
equations. 
If you use this ansatz to calculate the various
coefficients, you find that in both cases, {\it all} coefficients are
proportional to $v_0^{\pm}$. For the radiation dominated case, 
the coefficients are a finite series
in powers of $\tilde h$ ; for matter dominated, 
they are all proportional to $h'$. The
answer for the angle of rotation of the polarization plane, to lowest
non-trivial order in $1/x$ (remember that $x=k\eta$) is given by,
\begin{equation}
\Delta \phi = |\arg {v_0^+} - \arg {v_0^-} + 2 tan^{-1}( {\tilde h} k / x)|~ 
for ~ RD~ \label{rde} \end{equation}
and
\begin{equation}
\Delta \phi = |\arg {v_0^+} - \arg {v_0^-} + 2 tan^{-1} ({h'} k^3 / 3 x^3)|~ 
for ~ MD~ \label{mde} \end{equation}

But, recalling that the angle of rotation must vanish in absence of our 
proposed interaction, we get $\arg {v_0^+} - \arg {v_0^-} = 0$. Note that  
{\it no} assumption is made about the dependence of the coefficients on
$h$. Thus, the answers for 
the angle of rotation are : 
\begin{equation}
\Delta \phi = |2 tan^{-1}( {\tilde h} / \eta)|~ 
for ~ RD~ \label{rdef} \end{equation}
and
\begin{equation}
\Delta \phi = | 2 tan^{-1} ({h'} / 3 \eta^3)|~ for ~ 
MD~ \label{mdef} \end{equation}

For very small $h$, the inverse tangent may be replaced by its argument. 
These are our predictions from theory (to the lowest order in $h$) 
for the rotation 
angle, which may now be checked against the data. It is more
convenient to rewrite the expressions in terms of the red-shift
$z$ -- these, i.e. the expression for the lookback time can be
obtained from the expression 

\begin{equation}
t-t_0 = \frac{2}{3H_0}\left [ 1- \left (1+z\right )^{-\frac{3}{2}}
\right ]
\end{equation}

where $H_0$ is the value of the Hubble parameter today. The
relation between conformal time $\eta$ and real time $t$ can
be easily obtained from the expression,  $a(\eta)d\eta = dt$.

One can obtain expressions for the rotation of the electric field
which to the lowest order turns out to be the same as for the
magnetic field. We also note the fact that at the lowest order
$\bf E.B$ is equal to zero, but it may not be so beyond this
order.

\section{Conclusions}

A part of the rotation of the plane of polarisation of light 
emitted from distant galaxies obtained after subtracting
out the Faraday component is claimed to be due to the presence of the
Kalb--Ramond field. The modified Maxwell equations after the
inclusion of the effects of the H field have been written and
analysed both in flat and curved FRW backgrounds. The wave 
equations for the electric and magnetic fields are indeed
different (unlike usual electromagnetism) and the effects on the
rotation are generally different. 

We have calculated the rotation for three different cases both for the
electric and magnetic fields. The results are as follows.

(i) For a flat spacetime, $\left (\Delta \phi \right )_{mag}
=-h\eta$ (here $\eta$ and the real time $t$ are actually the
same). $\left ( \Delta \phi\right )_{elec}$ may be the same 
or different. We are able to write down $a$ solution for which
the electric and magnetic field rotations are the same except for
a phase of $\frac{\pi}{2}$ between the two. 
 
(ii) For the realistic scenarios , an asymptotic series solution valid
in the large $\eta$ regime is obtained for both the electric and
magnetic fields. The rotation for the radiation and matter dominated
cases go as $\frac{1}{\eta}$ and $\frac{1}{\eta^3}$ respectively
. It is possible to use the expressions for 
conformal time in terms of the redshift and demonstrate the rotation
explicitly in terms of the redshift.

In regard to astrophysical observations of the optical activity we have
discussed, our primary motivation is the careful analysis by Jain and Ralston
\cite{jr}, which is free of contentious issues pertaining to the acquisition and
analysis of data. Such issues were in focus after the incipient work of Nodland
and Ralston \cite{nr}, and appears to have been resolved satisfactorily. It is
fair to say therefore, that there is definite evidence that the rotation of the
plane of polarization of radiation travelling over cosmologically large
distances is not entirely attributable to the Faraday rotation due to magnetic
fields present in the galactic plasma. In this paper, we have presented 
arguments to the effect that the
`primordial' optical activity is quite likely due to a KR (axion)
field which endows the spacetime in its immediate vicinity with
torsion. It is perhaps not without significance that for both the
radiation and matter dominated scenarios, the calculated angle of
rotation of the plane of polarisation is independent of wavelength. This
property is shared by the angle $\chi$ which is the intercept vide
Eqn (1) of the straight line $\theta $ versus $\lambda^{2}$. 

All this is perhaps an indication that supergravity is at work, and what one 
is observing is perhaps a massless mode of an underlying string
theory, hitherto unobserved because of its weak coupling to other matter. As
pointed out in {\cite{carr}}, such a weak coupling could be detected perhaps in
future, if not through presently available data. We leave open the question
whether the present data does actually substantiate our theoretical conclusions.
 
Apart from the astrophysical ramifications of our work, the fact that the
only known proposal of coupling the Maxwell field to an Einstein-Cartan
geometry in a gauge invariant manner leads directly to the optical
activity discussed above can, in principle, be of significant use in the
detection of torsion as a geometrical property of spacetime. More
quantitative analysis of this aspect will form the subject of future
publications. 

\section*{Acknowledgements}

The work is supported by
Project grant no.98/37/16/BRNS cell/676 from The Board of Research in Nuclear
Sciences, Department Of Atomic Energy, Government of India.
We acknowledge useful correspondence with P. Jain. SK thanks 
S. Bharadwaj for useful discussions. AS acknowledges the local
hospitality provided by the Institute of Mathematical Sciences, Chennai, India
where part of this work was done. PM would like to
thank J. Ambjorn, A. Ashtekar, S. Chaudhuri and H. Nicolai for discussions, and
the Centre for Gravitational Physics, Penn State University, USA, The Niels Bohr
Institute, Copenhagen, Denmark and the Albert Einstein Institute, Potsdam,
Germany, for hospitality during the
completion of this work.

\end{document}